\documentstyle[11pt,newpasp,twoside,epsf]{article}

\def\edcomment#1{\iffalse\marginpar{\raggedright\sl#1\/}\else\relax\fi}
\reversemarginpar

\begin{document}

\title{The 1 Ms Chandra Survey of the \hbox{HDF-N}: Populations at the Faintest X-ray Fluxes}
\author{ANN E. HORNSCHEMEIER}
\affil{THE PENNSYLVANIA STATE UNIVERSITY, 525 DAVEY LAB, UNIVERSITY PARK, PA 16802}

\author{CDF-N TEAM}
\affil{PSU, UH IFA, UW-MADISON, MIT CSR, CMU, CALTECH}

\begin{abstract}
The Chandra Deep Field-North survey, which has at its center the Hubble Deep
Field-North, has reached an exposure of 1 Ms and is now available to
the public for analysis.  This great astronomical resource   
will soon be released to the community in the form of a catalog paper with
accurate X-ray fluxes in four bands and astrometry good to 
$\approx 0.6^{\prime \prime}$--$1.2^{\prime \prime}$ over the 
entire ACIS-I field.

The scientific focus of this contribution is the population of X-ray
sources detected at X-ray fluxes below the faintest detection limits of
X-ray observatories such as ROSAT and ASCA.  These include fairly
normal and star-forming galaxies out to $z\approx2$, starburst galaxies
from $z=2$--4 and possibly
very high redshift ($z> 6$) AGN.  The exciting new prospects
for studying these populations in the X-ray band are discussed.
\end{abstract}

\section{Introduction: The Chandra Deep Field-North Survey}

Our team is in the process of making an extremely deep survey of
the Hubble Deep Field North (hereafter \hbox{HDF-N}) and its environs with
the Advanced CCD Imaging Spectrometer
on board the {\it Chandra X-ray Observatory\/}. This is one of the two deepest X-ray
surveys ever performed; for point sources near the aim point it 
reaches 0.5--2.0~keV and 2--8~keV flux limits of 
$\approx 3\times 10^{-17}$~erg~cm$^{-2}$~s$^{-1}$ and
$\approx 2\times 10^{-16}$~erg~cm$^{-2}$~s$^{-1}$, respectively.  
Within the $\approx 20^{\prime} \times 22^{\prime}$ area (referred to
as the Chandra Deep Field-North region, hereafter the ``\hbox{CDF-N}"), 370 point sources and 
several extended sources are
detected in the ACIS data.  The 0.5--8~keV adaptively smoothed image of the
field is shown in Figure~1.   The detailed analysis of the \hbox{CDF-N} data is described in
the catalog paper Brandt et al. (2001b; hereafter referred to as Paper V). 

Figure~1 shows the outline of the \hbox{HDF-N} within the \hbox{CDF-N}, which comprises
about 2\% of the entire field.  There are 15 X-ray sources detected in the
this region (Paper V).  These sources span a wide range of properties, with 0.5--2~keV
rest frame luminosities between $\approx 2 \times 10^{39}$~erg~s$^{-1}$ and
$\approx 5 \times 10^{44}$~erg~s$^{-1}$, redshifts from $z=0.089$ to $z=3.479$,
and a range of X-ray hardness ratios.  

Figure~2 places the 1~Ms \hbox{CDF-N} survey into scientific perspective.  Shown
are a variety of 0.5--2~keV X-ray surveys (chosen to be representative, not complete!). 
The flattening of the number counts is apparent as
the net gain in number of sources per unit solid angle decreases at the faintest
fluxes.  One might find this flattening of the number counts discouraging,
but in fact these faintest fluxes are quite tantalizing, for it is here
that we reach the limits necessary to detect the most distant and most numerous 
classes of X-ray emitters. 

The most distant are perhaps 
extreme redshift ($z > 6$) quasars, a population which was emitting X-rays very
soon after the dawn of the modern Universe.
These extremely distant objects are expected
to be fairly optically faint, and studying them was one of the motivating factors
for a comprehensive examination of optically faint ($I > 24$) sources within
the \hbox{CDF-N} 1~Ms survey by Alexander et al. 2001 (hereafter Paper VI).  A fair
fraction of the X-ray background (15--30\%) at the fainter flux levels is
 accounted for by these optically faint sources, and it is in this region of
parameter space that one would expect $z > 6$ quasars to live. 

The most numerous are ``normal" 
galaxies where 
X-ray binaries, hot ISM, and supernovae contribute as much or more to the X-ray
luminosity as accretion onto a nuclear black hole (Hornschemeier et al. 2001, 
hereafter Paper~II and references therein; Brandt et al. 2001a, hereafter Paper~IV).
These sources typically have X-ray luminosities of $10^{39}$--$10^{41}$~erg~s$^{-1}$,
which can be reached to $z\approx1$ in $\approx 1$~Ms 
{\it Chandra} observations.  It is predicted through modelling and
through statistical analysis of the \hbox{CDF-N} data that the typical spiral 
galaxy will be detected in the 0.5--2~keV band at flux levels of $\approx 5 \times 10^{-18}$~erg~cm$^{-2}$~s$^{-1}$
(Ptak et al. 2001; A. E. Hornschemeier et al., in prep; hereafter Paper~VIII) and 
that normal galaxies will account for $\approx 5$\% of the soft X-ray background.

In this contribution to the ASP Sharp Focus proceedings, I discuss 
both the optically faint (possibly $z > 6$) and the optically 
bright (low $z$ normal galaxies) classes of X-ray sources, 
the reader is encouraged to consult the relevant \hbox{CDF-N} papers for more details.  

\begin{figure}
\caption{Adaptively smoothed image of the \hbox{CDF-N} 
in the 0.5--8~keV band.  
This image has been binned by a factor of four in both right ascension and declination. 
The adaptive smoothing has been performed using the code of 
Ebeling, White, \& Rangarajan (2001) at the $2.5\sigma$ level. 
Much of the apparent diffuse emission
is instrumental background.
The light colored grooves running through the 
images correspond to the gaps between the CCDs. The small polygon indicates 
the \hbox{HDF-N} itself.}
\vspace{2.5truein}
\plotfiddle{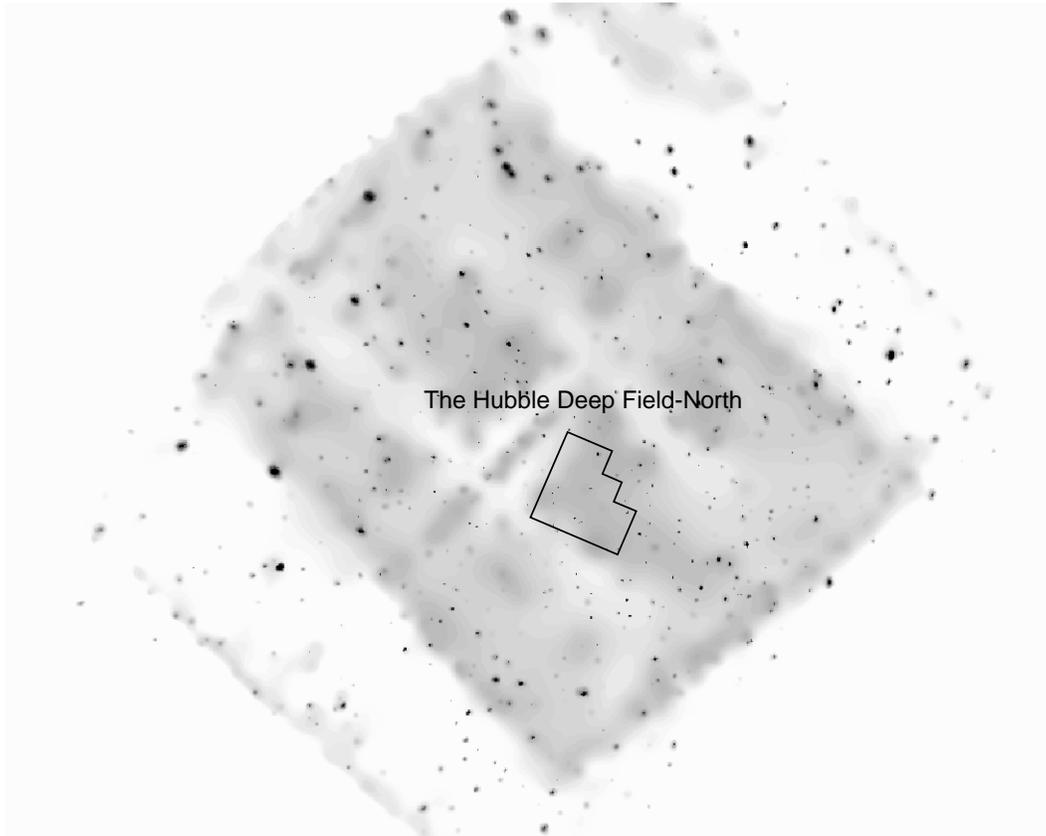}{5.5in}{0.0}{75}{75}{-225.0}{50.0}
\end{figure}

\begin{figure}
\caption{Distribution of some extragalactic X-ray surveys in the 0.5--2.0~keV 
flux limit versus solid angle plane, adapted from Paper~V. Shown are 
the {\it Uhuru} survey (e.g., Forman et~al. 1978), 
the {\it ROSAT} All-Sky Survey (RASS; e.g., Voges et~al. 1999), 
the {\it Einstein} Extended Medium-Sensitivity Survey (EMSS; e.g., Gioia et~al. 1990), 
the {\it ROSAT} International X-ray/Optical Survey (RIXOS; e.g., Mason et~al. 2000), 
the {\it XMM} Serendipitious Surveys ({\it XMM} Bright, {\it XMM} Medium, {\it XMM} Faint; e.g., Watson et~al. 2001)
, 
the {\it Chandra} Multiwavelength Project (ChaMP; e.g., Wilkes et~al. 2001),
the  {\it ROSAT} Ultra Deep Survey (UDS; e.g., Lehmann et~al. 2001), 
the deep survey of the Lockman Hole ( {\it XMM} LH; e.g., Hasinger et~al. 2001), 
{\it Chandra} 100~ks surveys (e.g., Mushotzky et~al. 2000), and 
{\it Chandra} 1~Ms surveys (i.e., the current one and the {\it Chandra} Deep 
Field South survey, see Colin Norman's CDF-S contribution to these proceedings)
Solid dots are for surveys that have been completed, and open circles
are for surveys that are in progress. 
The dotted curves show, from top to bottom, the loci of 100, 1000, and 
10000 0.5--2.0~keV sources (these have been calculated using the
number counts of Hasinger et~al. 1998 and Garmire et al. (2001; Paper III); 
for example, a
1~degree$^2$ survey with a 0.5--2.0~keV flux limit of 
$9.5\times 10^{-15}$~erg~cm$^{-2}$~s$^{-1}$ will detect
$\approx 100$ sources. 
}
\vspace{2.5truein}
\plotfiddle{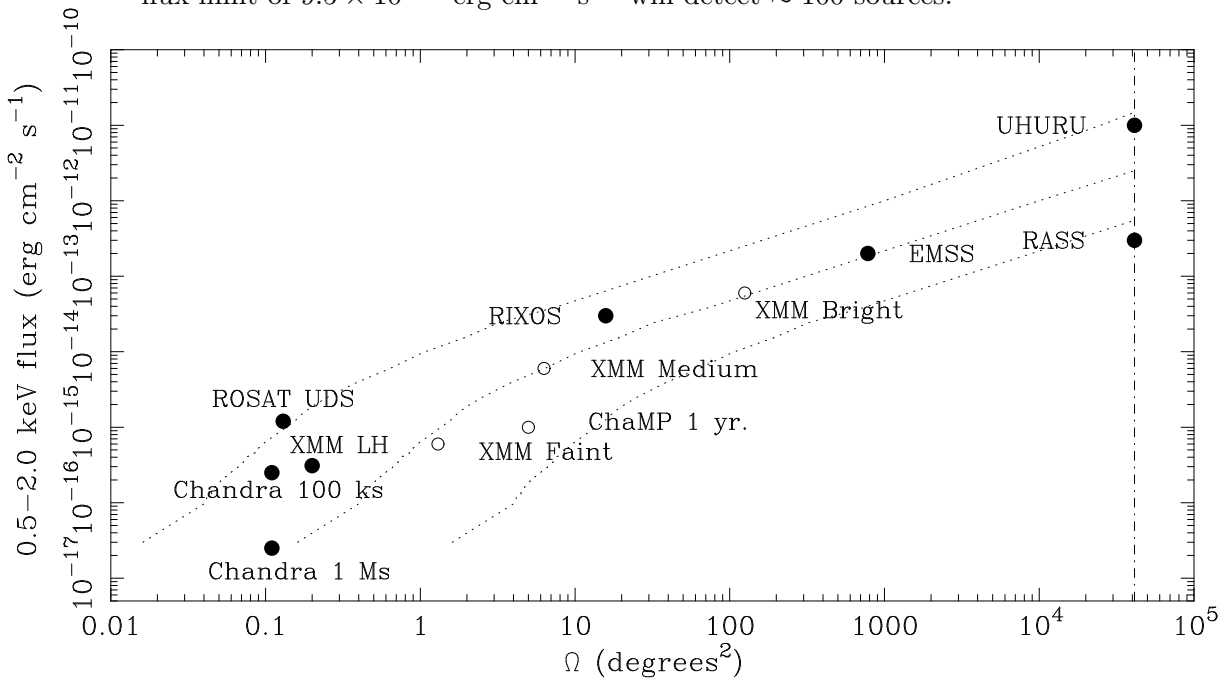}{1.5in}{270.0}{75}{75}{-300.0}{450.0}
\end{figure}

Neither of these source populations (optically faint sources and ``normal"
galaxies) are particularly unexpected from an observational point of view.
Most of the X-ray background
flux is explained by active galaxies, which have been observed to typically
exist within a well-defined range of X-ray-to-optical flux ratios (Schmidt
et al. 1998).  Thus, if one extends this trend to faint X-ray fluxes, the
optical counterparts predicted are indeed quite optically faint (see Figure~3).
Below 0.5--2~keV fluxes of $\approx 1 \times 10^{-15}$ erg~cm$^{-2}$~s$^{-1}$,
a source population is seen to arise with X-ray-to-optical flux ratios lower
than typically seen in AGN populations, this is simply due to the flux limits being sufficiently low to detect non-AGN activity at low-to-moderate redshift ($z < 1$).

\begin{figure}
\caption{Plot of $R$ magnitude versus 0.5--2~keV flux for X-ray 
detected sources, adapted from Paper II. Solid triangles are sources 
from the \hbox{CDF-N} data of Paper~II,  and
open circles are those from Mushotzky et~al. (2000).
 Solid dots are sources
from Schmidt et~al. (1998). 
We plot only the AGN from Schmidt et~al. (1998).
The two stars are spectroscopically identified stars
The vertical dotted lines show the Paper~II  
 detection limits for an assumed $\Gamma=1.4$ power-law
spectrum; sources slightly beyond these lines have spectral
shapes differing from a $\Gamma=1.4$ power law.  The boxes mark our sources with
extreme $\log{({{f_{\rm X}}\over{f_{\rm R}}})}$ values.}
\vspace{2.5truein}
\plotfiddle{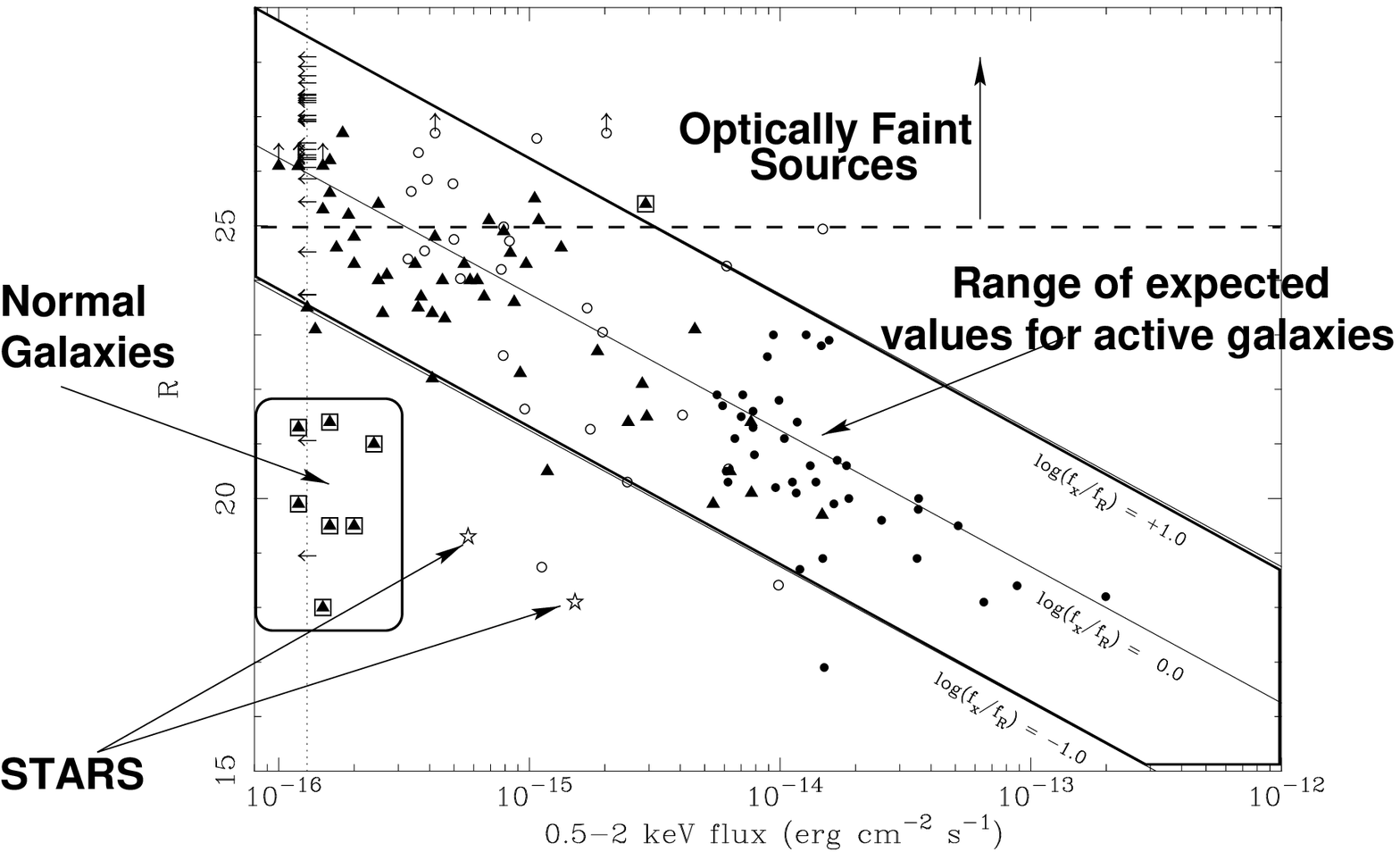}{2.5in}{0.0}{85}{85}{-250.0}{50.0}
\end{figure}

\section{Optically Faint ($I > 24$) X-ray Sources}

A detailed analysis of optically faint ($I\ge24$) X-ray sources 
identified within the 1~Ms \hbox{CDF-N} exposure has been made in Paper~VI.  In this paper,
we studied an $8.4^{\prime}\times8.4^{\prime}$ region within the Hawaii flanking-field area
 containing the Hubble Deep Field North region.   Analysis of the optical and X-ray 
properties suggests a large number of these optically faint X-ray sources are 
likely to host obscured AGN activity at $z=$~1--3 (Paper VI).   

The X-ray flux distribution of the optically faint sources is consistent with that
of the optically bright source population, indicating that this population does
not suddenly arise below some faint X-ray flux level.  Their X-ray-to-optical flux
ratios are consistent with AGN in the local Universe (see Figure~3).  Their fairly
red optical colors are not consistent with these objects being unobscured broad-line
 AGN and, as a population, they have flatter X-ray slopes on average than the optically
bright sources (Paper~VI).  Note that while {\it on average} they are X-ray harder,
they are not all X-ray hard; there is at least one apparently normal broad-line AGN within
the optically faint \hbox{CDF-N} population studied in Paper~VI.  However, the majority
of the optically faint X-ray sources are consistent with being obscured AGN (Paper~VI).

Obscured AGN have their optical emission dominated in most cases by the
host galaxy and this property can be used to place constraints on source
redshifts. In Figure~4 (adapted from Paper~VI), SED
tracks of $I$-band magnitude vs. redshift are plotted, showing the way
in which different classes of objects become fainter at increasing
redshift.  We expect that obscured AGN and fairly normal galaxies will
follow the SED tracks for the Sc, Sa, or E type galaxies rather than the
QSO SED.  Overplotted are the data showing the average spectroscopic
redshifts for the optically bright population, excluding the confirmed
broad-line AGN: these X-ray sources do indeed follow the SED tracks of
normal galaxies.  One can then extend these tracks to fainter optical
magnitudes to place some constraints on their redshifts.  It is found
that for $I > 24$, one expects the objects to have $z=$~1--3.  For the
few optically faint sources with either photometric or spectroscopic
redshifts, this appears to hold reasonably well (Paper~VI).

\begin{figure}
\caption{$I$-band vs. $z$ tracks for different SEDs with \hbox{CDF-N} data overplotted,
(adapted from Paper~VI). 
The filled triangles are the broad-line AGN, the filled
squares are the luminous narrow-line AGN, the open circles are optically faint
X-ray sources with spectroscopic or photometric redshifts.  
The filled circles are the average spectroscopic
redshifts for the $I<23$ {\it Chandra} sources.
The crosses are the average photometric redshifts for optical sources in the \hbox{HDF-N}
(from Fern\'andez-Soto, Lanzetta, \& Yahil 1999). The solid,
long-dashed and short-dashed curves are the redshift tracks of
$M_{I}=-23$ E, Sa and Sc host galaxies. The dotted curve is the
redshift track of an $M_{I}=-23$ QSO (for details on these calculations,
see Paper~VI). This figure suggests
that if the optically faint X-ray sources are the high-redshift
analogs of the optically bright X-ray sources, the majority should lie
at $z\approx$~1--3.}
\vspace{2.5truein}
\plotfiddle{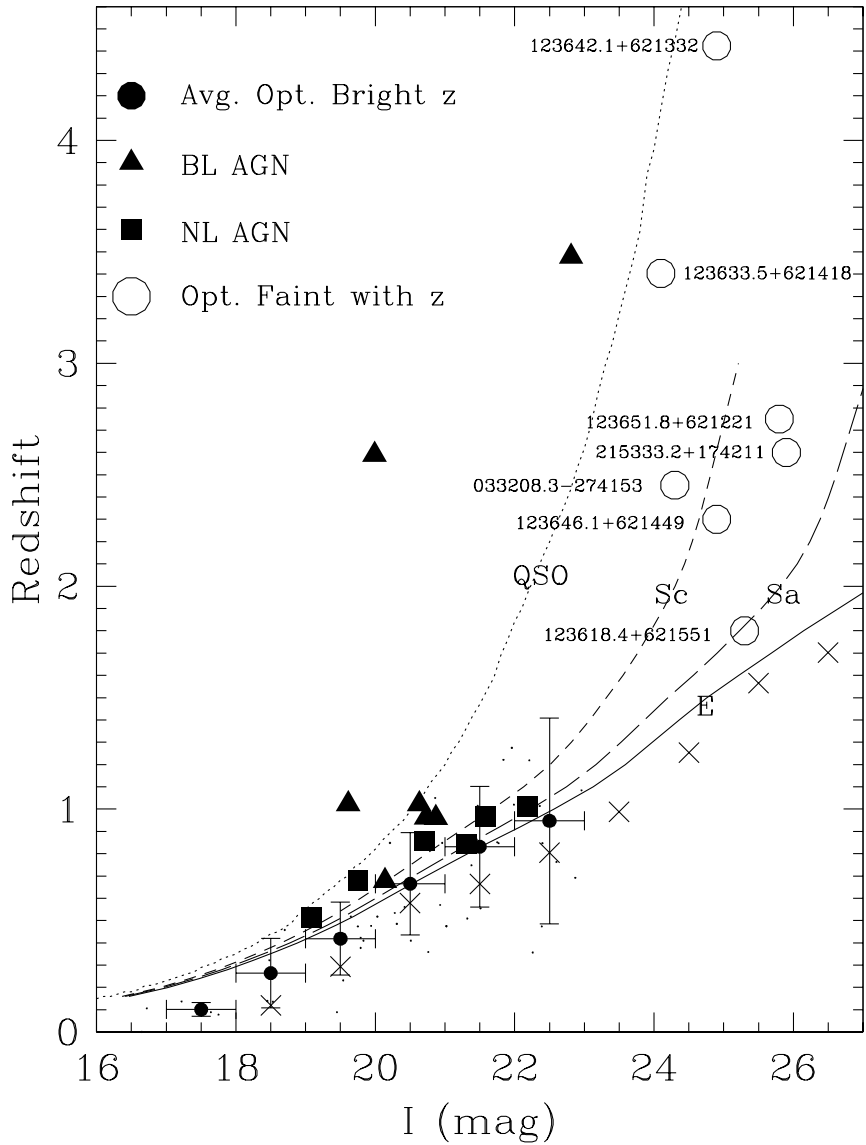}{4.5in}{0.0}{60}{60}{-200.0}{50.0}
\end{figure}

Extreme redshift QSOs ($z > 6$) would not only be  optically faint but
``optically blank" even in the very deep ($I \approx 25.3$) optical
imaging data of the \hbox{CDF-N} due to the Lyman break entering the
$I$-band at $z \approx 6$.   Within the area under study in Paper~VI,
there were only 15 sources having no counterparts down to $I\approx
25.3$ ($\approx 10$\% of the sources). 
Based on a simple hierarchial
cold dark matter model and using constraints from the QSO X-ray
luminosity function, Haiman \& Loeb (1999) predicted $\approx 15$ QSOs
(i.e.,\ $L_X>10^{44}$ erg s$^{-1}$) at $z > 6$ at the depth and area
of our survey.  While this is comparable to the number of optically
blank sources detected, for several reasons described in Paper~VI, it
is likely that most of the optically blank X-ray sources lie at similar  
redshifts to the optically faint X-ray sources and hence at $z<6$.  We
are thus placing the first significant constraints on this population and
are currently not finding evidence for a large population of extremely high 
redshift AGN.

\section{Optically Bright X-ray Sources: Constraints on Normal Galaxies}

\subsection{Galaxies at \boldmath$0.3 < z < 2.0$: Probing the X-ray Response to the Star-Formation Peak}

The \hbox{CDF-N} has reached the depths necessary to detect fairly normal 
galaxies to $z\approx0.3$ (see Figure~5 for an example), or look-back
times of several Gyr.  Although these normal galaxies comprise only 10--15\% of the soft-band sources in
the \hbox{CDF-N} survey, it is expected that they will dominate the number counts 
at 0.5--2~keV fluxes of
$\approx 1 \times 10^{-17}$--$1 \times 10^{-18}$ erg~cm$^{-2}$~s$^{-1}$
(Ptak et~al. 2001).   The emission mechanisms in the galaxies detected thus
far are quite diverse:  several are dominated by ultra-luminous off-nuclear
X-ray sources which are most likely binaries, other show signs of being 
low-luminosity AGN, whereas the typical galaxy has an X-ray hardness ratio consistent
with binary populations (Paper II; Tozzi et al. 2001; 
A.E. Hornschemeier et al., in preparation).

\begin{figure}
\caption{An X-ray detected normal galaxy at $z\approx 0.1$ in the \hbox{CDF-N} (A.E. Hornschemeier et al.,
in preparation). 
The X-ray source is $\approx 3.0^{\prime \prime}$ from the galaxy's nucleus and
has an X-ray luminosity of $\approx 2 \times 10^{39}$~erg~s$^{-1}$.  It is 
coincident with a knot on one of the spiral arms of the galaxy and is most likely
a super-Eddington X-ray binary system or starburst region.  This cut-out is
from a deep
$V$ image that covers the entire \hbox{CDF-N} field (see A. Barger et~al., in 
preparation for more information on the deep optical image).
}
\vspace{2.5truein}
\plotfiddle{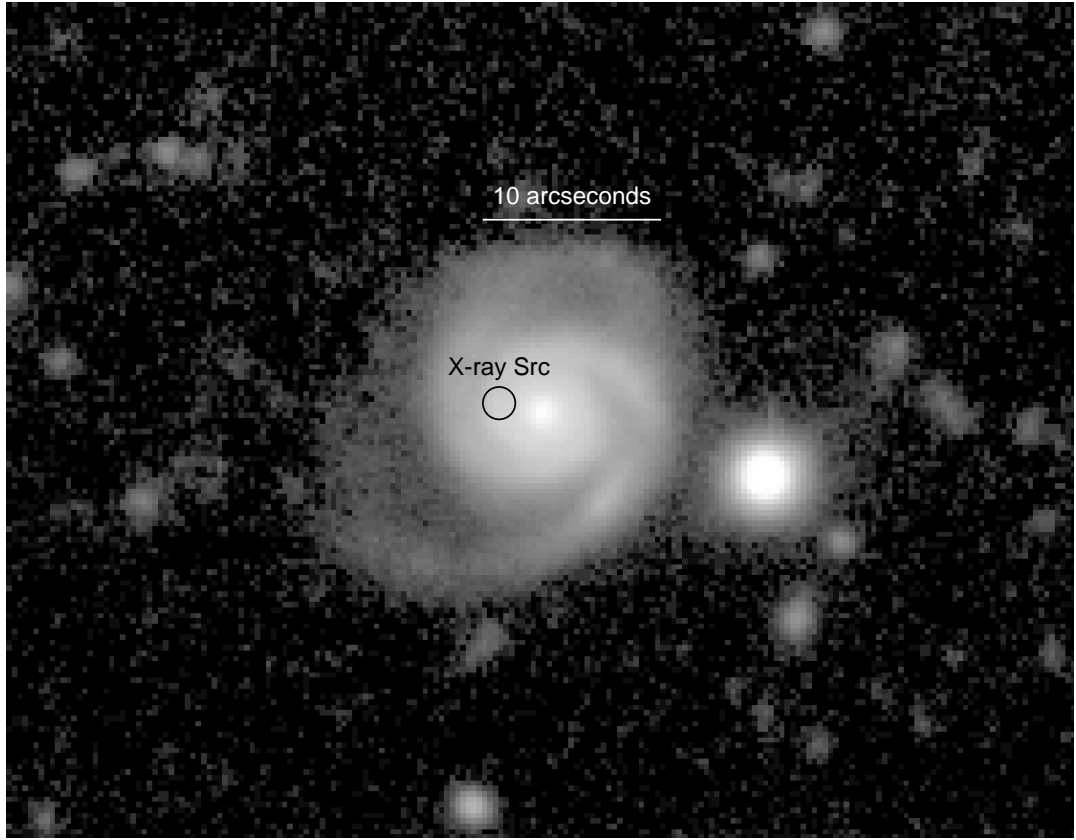}{4.5in}{0.0}{75}{75}{-200.0}{50.0}
\end{figure}

It is expected that the X-ray luminosity per unit $B$-band luminosity for normal
galaxies will increase 
at $z\approx 0.5$--1 
due to the production of X-ray binaries from $z\approx1$--3 (e.g., Ghosh \& White 2001;
Ptak et al. 2001).
It is possible to test this expectation despite the 
anticipated extremely faint X-ray flux of the typical galaxy by stacking the
data from individually undetected galaxies.
We have empirically assessed the stacking false-detection probabilities by performing
Monte-Carlo simulations designed to reproduce the 
actual stacking as closely as possible.  Figure~6 shows an example
of 100,000 such trials, the resulting
distribution is closely Gaussian and this particular stacking result 
was expected for less than 1 of those 100,000 trials, indicating a highly
significant (i.e., 99.999\% confidence) average detection (see Paper~IV for more details).

\begin{figure}
\caption{Results from the Monte-Carlo testing of the galaxy stacking analysis. 
We performed 100,000 stacking trials at randomly 
selected positions, and plotted the number of trials yielding a given number 
of counts in the stacking aperture (see Paper~IV for a more detailed 
description).
The arrow indicates the number of counts obtained for the galaxies of interest;
the detection is significant at the $> 99.99$\% level.
}
\vspace{2.5truein}
\plotfiddle{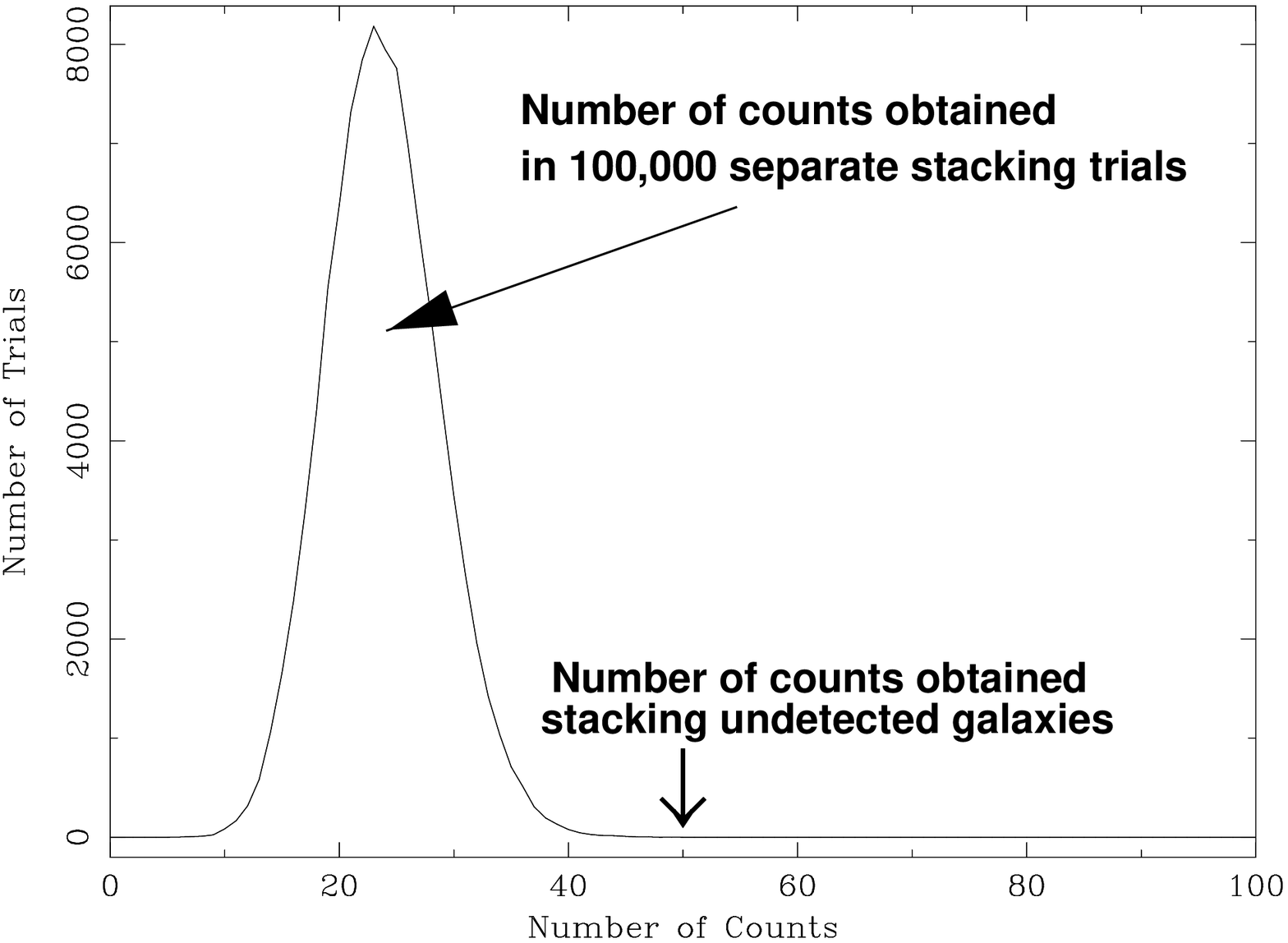}{4.5in}{0.0}{75}{75}{-200.0}{50.0}
\end{figure}

These stacking analyses have allowed us to study samples of spiral galaxies
within the \hbox{HDF-N} and its environs;  we have found the typical spiral galaxy
is detected at 0.5--2~keV X-ray fluxes of 
$\approx$(3--5)$\times 10^{-18}$~erg~cm$^{-2}$~s$^{-1}$ (Paper VIII).
The X-ray to $B$-band luminosity ratios,
which indicate the level of current and recent star-formation activity in a normal
galaxy, are found to not evolve upwards by more than a factor of $\approx 2$ to
$z\approx 1$ (Paper~IV, Paper VIII) but some upwards evolution is detected
by $z\approx 2$.  Since different global star-formation rates can lead to 
very different X-ray luminosity evolution profiles (e.g. Ghosh \& White 2001),
these constraints on evolution of galaxies in the X-ray band are a useful
independent probe of the cosmic star-formation history.  For more details, please
see Paper~VIII.

\subsection{Galaxies at \boldmath$2 < z < 4$: Lyman Break Galaxies}

Over the last few years, the Lyman break technique has been used extensively to
isolate galaxies at $z\approx$~2--4 (e.g., Steidel et~al. 1996; 
Lowenthal et~al. 1997; Dickinson 1998) and hence observe activity
 near the peak of the cosmic star-formation rate (e.g., Blain et~al. 1999). 
Lyman break galaxies often exhibit stellar and interstellar absorption lines
characteristic of local starburst galaxies, and have
$B$-band luminosities somewhat larger than present-day
$L^\ast$. Their morphologies are varied, with multiple knots of emission
and diffuse wispy tails that suggest nonrelaxed systems. 

We selected galaxies within the \hbox{HDF-N} with $z=$2--4 for stacking analysis using the 
available spectroscopic redshift catalogs (e.g. Cohen et al. 2000; see 
references in Paper~VII);  there are 28 such objects, and most 
were found using the Lyman break technique (the {\it HST} 
$U_{300}$ filter has allowed Lyman break galaxies to be 
found down to $z\approx 2$ in the \hbox{HDF-N}; e.g., Dickinson 1998).  
The stacked images have effective exposure times of 
22.4~Ms (260 days) and the average rest-frame 2--8~keV luminosity 
of a Lyman break galaxy is derived to be $\approx 3.2\times 10^{41}$~erg~s$^{-1}$,
comparable to that of the most X-ray luminous starbursts in the local Universe.
The observed ratio of X-ray to $B$-band luminosity is consistent 
with that seen from local starbursts. The X-ray emission probably 
arises from a combination of high-mass X-ray binaries, ``super-Eddington''
X-ray sources, and low-luminosity active galactic nuclei (see Paper VII).

\section{The Future}

Members of the \hbox{CDF-N} team (A. Barger and L. Cowie) are obtaining 
extremely deep wide-field optical and near-infrared coverage and will 
be able to further the analysis performed in Paper~VI.
The optically faint X-ray source population should be better constrained 
by using the properties of the multi-band optical photometry to obtain photometric redshifts.

The stacking results show that {\it Chandra} ACIS performs well at 
source detection even with effective exposure times of 260 days.
Any systematic effects
that cause the sensitivity to deviate from that expected by
photon statistics appear mild. Stacking analyses 
using deeper observations with {\it Chandra} will allow this 
work to be extended.  

The \hbox{CDF-N} survey will go to 2~Ms of coverage
over the course of the next year and the ultimate goal is to obtain 5~Ms of coverage
over the next $\approx 5$ years. This is an ambitious project and
will fulfill one of {\it Chandra's\/} central design goals and become a
long-lasting ($\approx 20$ year) legacy of {\it Chandra}, laying
the groundwork for the next generation of X-ray telescopes.  
{\it XEUS} and {\it Generation-X} are designed to detect
sources to $\sim$$10^{-18}$~erg~cm$^{-2}$~s$^{-1}$ and perform X-ray spectroscopy on sources
to $\sim$$10^{-17}$~erg~cm$^{-2}$~s$^{-1}$. A 5~Ms {\it Chandra}
observation will detect sources to
$\approx 10^{-17}$~erg~cm$^{-2}$~s$^{-1}$.  These observations will
bolster the scientific cases for these missions, not due to be operational for 15--20 years,
and will be invaluable when planning their design.

\end{document}